\documentclass[doublecol]{epl2} 
    
\newcommand{\Jnature}{Nature (London)}
\newcommand{\Jnatphys}{Nature Phys.}

\newcommand{\Jscience}{Science}

\newcommand{\Jprl}{Phys. Rev. Lett.}

\newcommand{\Jpra}{Phys. Rev. A}
\newcommand{\Jprb}{Phys. Rev. B}

\newcommand{\Jpre}{Phys. Rev. E}

\newcommand{\Jnjp}{New J. Phys.}
\newcommand{\Jepjd}{Eur. Phys. J. D}

\newcommand{\JApplPhysLett}{Appl. Phys. Lett.}

\newcommand{\Jjetp}{Sov. Phys. JETP}

\usepackage[english]{babel}
\usepackage[latin1]{inputenc}
\usepackage{latexsym}
\usepackage{graphics}
\usepackage{subfigure}
\usepackage{epsfig}
\usepackage{color}
\usepackage{hyperref}
\usepackage[normalem]{ulem}  
\hypersetup{
colorlinks=true,
citecolor=blue,
linkcolor=red,
urlcolor=black
}

\newcommand{\ie}{{i.e.}}
\newcommand{\eg}{{e.g.}}
\renewcommand{\etal}{{\textit{et al.}}}

\renewcommand{\vect}[1]{\mathbf{#1}}
\newcommand{\av}[1]{\overline{#1}}
\newcommand{\geomav}[1]{#1^{\textrm{av}}}

\newcommand{\kB}{k_\textrm{\tiny B}}

\newcommand{\Vr}{V_\textrm{\tiny R}}

\newcommand{\sigmaOrth}{\sigma_{\perp}}
\newcommand{\sigmaPara}{\sigma_{\parallel}}

\newcommand{\Cor}{C}
\newcommand{\cor}{c}
\newcommand{\TFCor}{\tilde{C}}
\newcommand{\TFcor}{\tilde{c}}

\newcommand{\kE}{k_\textrm{\tiny \textit E}}

\newcommand{\DB}{D_\textrm{\tiny B}}

\newcommand{\DiffTensB}{\vect{D}_{\textrm{B}}}
\newcommand{\DiffTens}{\vect{D}_{*}}
\newcommand{\DiffTensCor}{\Delta\vect{D}}

\newcommand{\LocTens}{\vect{L}_\textrm{loc}}
\newcommand{\LocTensTwo}{\vect{\Lambda}}

\newcommand{\Emob}{E_{\textrm{c}}}

\newcommand{\EmobNoShift}{E^\prime_{\textrm{c}}}

\newcommand{\ud}{\mathrm{d}}
\newcommand{\vecr}{\textbf{r}}
\newcommand{\veck}{\textbf{k}}
\newcommand{\uveck}{\hat{\textbf{k}}}
\newcommand{\uvecx}{\hat{\textbf{x}}}
\newcommand{\uvecy}{\hat{\textbf{y}}}
\newcommand{\uvecz}{\hat{\textbf{z}}}
\newcommand{\uvecX}{\hat{\textbf{X}}}
\newcommand{\uvecY}{\hat{\textbf{Y}}}
\newcommand{\uvecZ}{\hat{\textbf{Z}}}
\newcommand{\vecq}{\textbf{q}}

\newcommand{\be}{\begin{equation}}
\newcommand{\beq}{\begin{eqnarray}}
\newcommand{\ee}{\end{equation}}
\newcommand{\eeq}{\end{eqnarray}}

\newcommand{\Gr}{G}
\newcommand{\Ga}{G^{\dagger}}

\newcommand{\eps}[1]{\epsilon(#1)}

\usepackage{amsmath}
\usepackage{amssymb}
\usepackage{graphicx}
\usepackage{color}

\newcommand{\remove}[1]{}

\title{Matter wave transport and Anderson localization in anisotropic three-dimensional disorder}
\shorttitle{Matter wave transport and Anderson localization in anisotropic three-dimensional disorder}

\author{Marie~Piraud \and Luca~Pezz\'e\thanks{Present address: INO-CNR and
LENS, Largo Enrico Fermi 6, I-50125 Firenze (Italy)} \and Laurent~Sanchez-Palencia\thanks{E-mail: \email{lsp@institutoptique.fr}}}
\shortauthor{M.~Piraud~\etal}

\institute{
  Laboratoire Charles Fabry,
  Institut d'Optique, CNRS, Univ Paris Sud,
  2 avenue Augustin Fresnel,
  F-91127 Palaiseau cedex, France
}


\abstract{We study quantum transport of matter waves in anisotropic three-dimensional disorder.
First, we show that structured correlations can induce rich effects, such as
  anisotropic suppression of the white-noise limit
  and inversion of the transport anisotropy.
Second, we show that the localization threshold (mobility edge) is strongly affected by a disorder-induced shift of the energy states, which we calculate.
Our work is directly relevant to ultracold-matter waves in optical disorder, and implications on recent experiments are discussed. It also offers scope for further studies of anisotropy effects in other systems with controlled disorder, where counterparts of the discussed effects can be expected.}

\begin{document}

\maketitle

\section{Introduction}
Coherent transport in disordered media is strongly affected by anisotropy effects.
It occurs in many systems, \eg\
  electrons in MOSFETs~\cite{bishop1984},
  diffusing-wave spectroscopy~\cite{pine1988},
  biomedical imaging~\cite{nickell2000}, as well as
  light in liquid crystals~\cite{kao1996,wiersma1999}, in phosphides~\cite{johnson2002}, or
    in microcavities~\cite{gurioli2005}.
So far, theoretical analyses mainly focused on models of disorder made of
  isotropic impurities imbedded in anisotropic media~\cite{woelfe1984,kaas2008} or
  stretched scatterers in isotropic media~\cite{tiggelen1996,stark1997},
which fairly describe the above systems.
The recent advent of systems where the disorder correlations can be controlled, \eg\
  tunable arrangements of scatterers for microwaves~\cite{kuhl2000,kuhl2008},
  engineered optical materials~\cite{barthelemy2008}
or
  ultracold atoms in optical disorder~\cite{lsp2010,clement2006},
opens new perspectives, and it becomes increasingly important to better understand
transport in disordered media with more complex correlations.

In this letter, we study quantum transport and Anderson localization (AL) of matter waves in three-dimensional (3D) disorder with structured, anisotropic correlations.
We first determine the diffusion and localization tensors using the self-consistent approach of ref.~\cite{woelfe1984}.
We show that appropriate correlations can lead to rich transport properties, such as
  anisotropic suppression of the white-noise limit and
  inversion of the transport anisotropy.
Then, going beyond the standard on-shell approximation of ref.~\cite{woelfe1984}, we
include the real part of the particle self energy.
While the latter affects the above results only quantitatively, we show that it strongly modifies
the behavior of the mobility edge.
Our results have direct implications to recent experiments with ultracold atoms in optical disorder~\cite{kondov2011,jendrzejewski2011}, which we discuss.
They can also be extended to waves with different dispersion relations and other models of disorder, where counterpart effects can be expected.

\section{Quantum transport theory}
Let us start with a brief theoretical reminder.
The building block to describe the density (intensity) propagation of a wave in a disordered medium is the four-point vertex
$ \Phi_{\veck,\veck'}(\vecq,\omega, E) \equiv 
\av{ \langle \veck_+ \vert \Gr(E_+) \vert \veck'_+ \rangle 
\langle \veck'_- \vert \Ga(E_-) \vert \veck_- \rangle}$
with $\Gr$ the retarded Green operator,
$\veck_\pm \equiv \veck \pm \vecq/2$ and $\veck_\pm' \equiv \veck' \pm \vecq/2$
the left and right entries,
$E_\pm \equiv E \pm \hbar\omega/2$,
and $(\vecq$, $\omega)$ the Fourier conjugates of the space and time variables~\footnote{Here, we use $\tilde{f}(\vecq,\omega) \equiv \int \ud\vecr \ud t\ f(\vecr,t)
  \exp[-i (\vecq\cdot\vecr - \omega t)]$.}.
The vertex $\Phi$ is governed by the Bethe-Salpeter equation, which can be formally written~\cite{rammer1998}
\be \label{BSE}
\Phi = \av{\Gr} \otimes \av{\Ga} + \av{\Gr} \otimes \av{\Ga} \, \mathrm{U} \, \Phi \,.
\ee
The first term in eq.~(\ref{BSE}) describes uncorrelated propagation of the field and its conjugate in the disordered medium. The second term
involves the vertex function $\mathrm{U}$, which includes all irreducible scattering diagrams, and
accounts for all correlations in the density propagation.
In the independent scattering (Boltzmann) and weak disorder (Born) approximations~\cite{woelfe1984},
$\mathrm{U}_{\veck,\veck'} (\vecq, \omega, E) \simeq  \TFCor (\veck - \veck')$,
where $\TFCor(\veck)$ is the disorder power spectrum (Fourier transform of
the correlation function~(see footnote$^1$)).
Here we choose the zero of energies such that the disorder is of zero average, \ie\ $\av{V}=0$.
For any other choice of the energy reference all energies appearing below should be shifted by $\av{V}$, \ie\ replace $E$ by $ E-\av{V}$.
At this stage, only the ladder diagrams in eq.~(\ref{BSE}) are retained. It represents an infinite series of independent scattering events, which leads to Drude-like diffusion.
The solution of eq.~(\ref{BSE}) is then dominated in the long time ($\omega \rightarrow 0$) and large distance ($\vecq \rightarrow 0$) limit by a diffusion pole~\cite{rammer1998}, which reads
\be \label{BSEsingular}
\Phi_{\veck,\veck'}(\vecq,\omega, E) 
= \frac{2\pi}{\hbar N_0(E)} \frac{\delta(E- \eps{\veck}) \, \delta(E- \eps{\veck'})}
{- i \omega + \vecq \! \cdot \! \DiffTensB(E) \! \cdot \! \vecq}
\ee
in the on-shell approximation [such that $\eps{\veck}=\eps{\veck'}=E$, where 
$\eps{\veck}$ is the disorder-free dispersion relation] and with $N_0(E)$ the disorder-free density of states.
For matter waves in free space (that we consider here), $\epsilon(\veck)=\hbar^2\veck^2/2m$ with $m$ the atom mass.
The components of the Boltzmann diffusion tensor $\DiffTensB(E)$ are given by~\cite{woelfe1984}
\beq 
&& D_{\textrm{B}}^{i,j}(E) = 
\frac{1}{N_0(E)} 
\bigg\{  
\Big\langle \tau_{E,\uveck} \, \upsilon_i \, \upsilon_j \Big\rangle_{\veck\vert E} 
\label{DBE} \\
&& + \frac{2 \pi}{\hbar } 
\sum_{\lambda^n_E \neq 1} \frac{\lambda^n_E}{1 \! - \! \lambda^n_E} 
\Big\langle \tau_{E,\uveck} {\upsilon}_i \phi_{E,\uveck}^n \Big\rangle_{\veck\vert E}
\Big\langle \tau_{E,\uveck} {\upsilon}_j \phi_{E,\uveck}^n \Big\rangle_{\veck\vert E}
\bigg\},
\nonumber
\eeq
where ${\upsilon}_i = \hbar k_i/m$ is the velocity along axis $i$,
$\tau_{E,\uveck} = \hbar/2\pi \langle \tilde{C}(\kE\uveck-\veck') \rangle_{\veck'\vert E}$
(with $\uveck \! \equiv \! \veck/\vert\veck\vert$ and $\kE \equiv \sqrt{2mE}/\hbar$)
is the on-shell scattering mean free time, and
$\langle ... \rangle_{\veck\vert E} \equiv \int \frac{\ud \veck}{(2 \pi)^3} \, ... \, \delta[E-\eps{\veck}]$
represents integration over the $\veck$-space shell defined by $\eps{\veck}=E$.
The functions $\phi_{E,\uveck}^n$ and the real-valued positive numbers $\lambda^n_E$,
indexed by $n$, are the solutions of the integral eigenproblem
\be \label{eigeq}
\frac{2\pi}{\hbar} \Big\langle \tau_{E,\uveck'} \tilde{C}(\kE\uveck-\veck') \phi_{E,\uveck'}^n 
\Big\rangle_{\veck'\vert E} = \lambda^n_E \, \phi_{E,\uveck}^n \,,
\ee 
normalized by
$\frac{2\pi}{\hbar} \big\langle \tau_{E,\uveck} \phi_{E,\uveck}^n\phi_{E,\uveck}^m \big\rangle_{\veck \vert E} = \delta_{n,m}$~\cite{woelfe1984}.
In the following, the above equations are solved numerically to determine $\DiffTensB(E)$ in anisotropic 3D models of disorder.

\section{Boltzmann diffusion in 3D speckles}
\begin{figure}[!t]
\begin{center}
\includegraphics[scale=0.35]{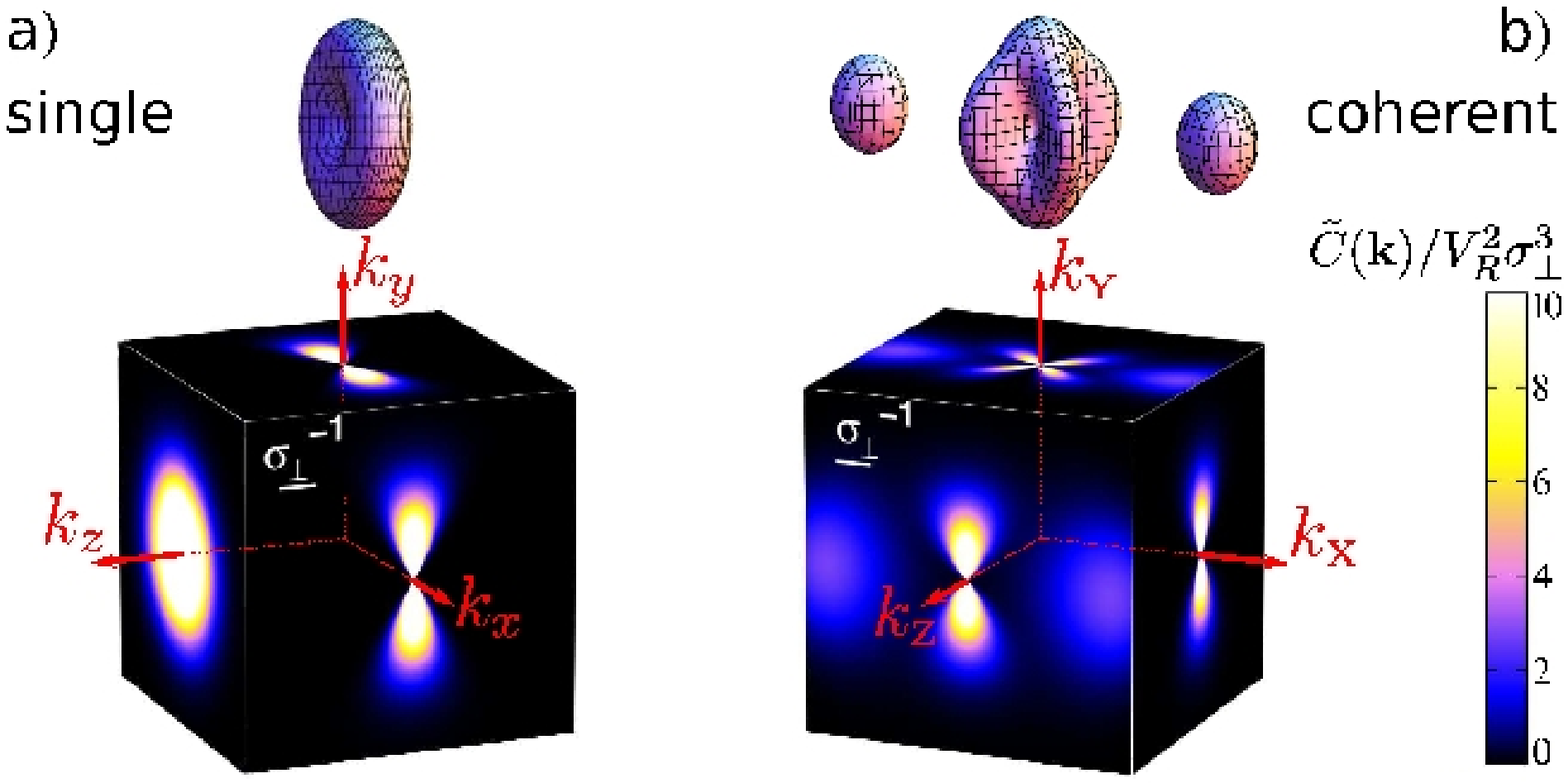}
\end{center}
\vspace{-0.3cm}
\caption{\small{(color online)} \label{fig:Speckle}
Disorder power spectrum $\TFCor(\veck)$ for the
  (a) single-speckle
  and (b) coherent-speckles
cases (Fourier transforms of the formulas given in the appendix) with
the parameters of refs.~\cite{kondov2011,jendrzejewski2011}
($\sigmaPara/\sigmaOrth\simeq 5.8$, and for (b) $\lambda_{\textrm{L}}/\sigmaOrth\simeq 2.16$).
The functions $\TFCor(\veck)$ are represented as iso-value surfaces (at $2\Vr^2\sigmaOrth^3$) and cuts in the planes defined by the transport eigenaxes (see text):
$\{\hat{\vect{u}}_x,\hat{\vect{u}}_y,\hat{\vect{u}}_z\}$ for (a)
and $\{\hat{\vect{u}}_X \equiv (\hat{\vect{u}}_x \! - \! \hat{\vect{u}}_z)/\sqrt{2}, \hat{\vect{u}}_Y \equiv \hat{\vect{u}}_y, \hat{\vect{u}}_Z \equiv (\hat{\vect{u}}_x \! + \! \hat{\vect{u}}_z)/\sqrt{2}\}$ for (b).
}
\end{figure}

Here we are particularly interested in the effects of structured correlations in 3D disorder.
In this respect, two configurations used recently for localizing ultracold atoms are well suited.
In ref.~\cite{kondov2011}
(\textit{single-speckle} case),
an optical speckle disorder is obtained using a single Gaussian laser beam of waist $w$ and wavelength $\lambda_{\textrm{L}}$, propagating along the $z$ axis, passed through a ground-glass plate and focused by a lens of focal distance $f$~\cite{goodman2007}. The disorder correlation function $C(\vecr)$ has correlation lengths $\sigmaPara={4 \lambda_{\textrm{L}} f^2}/{\pi w^2}$ along $z$ and $\sigmaOrth={\lambda_{\textrm{L}} f}/{\pi w}$ in the orthogonal plane ($x,y$) (see appendix). For $4f > w$, $\Cor (\vecr)$ is elongated along $z$.
Then, the disorder power spectrum $\TFCor(\veck)$ is isotropic in the ($k_x,k_y$) plane but significantly shorter along the $k_z$ axis and forms a plain-wheel shaped structure [see fig.~\ref{fig:Speckle}(a)].
In ref.~\cite{jendrzejewski2011} (\textit{coherent-speckles} case),
the disorder results from the interference of two mutually coherent and orthogonal speckle fields, propagating along the $z$ and $x$ axes, respectively. The power spectrum $\TFCor(\veck)$ then shows a complex structure, made of the sum of two orthogonally-oriented spectra similar to that of the single-speckle case, plus a coherence term~(see appendix and fig.~\ref{fig:Speckle}(b)).
Most importantly, the latter creates two broad structures (\textit{bumps}),
centered on the $\hat{\veck}_X \equiv (\hat{\veck}_x - \hat{\veck}_z)/\sqrt{2}$ axis.
In the following we call $\Vr \equiv \sqrt{C(\vecr=0)}$ the amplitude and $E_{\sigmaOrth} \equiv \hbar^2/m \sigmaOrth^2$ the correlation energy of the disorder.

\begin{figure}[!t]
\begin{center}
\includegraphics[scale=0.65]{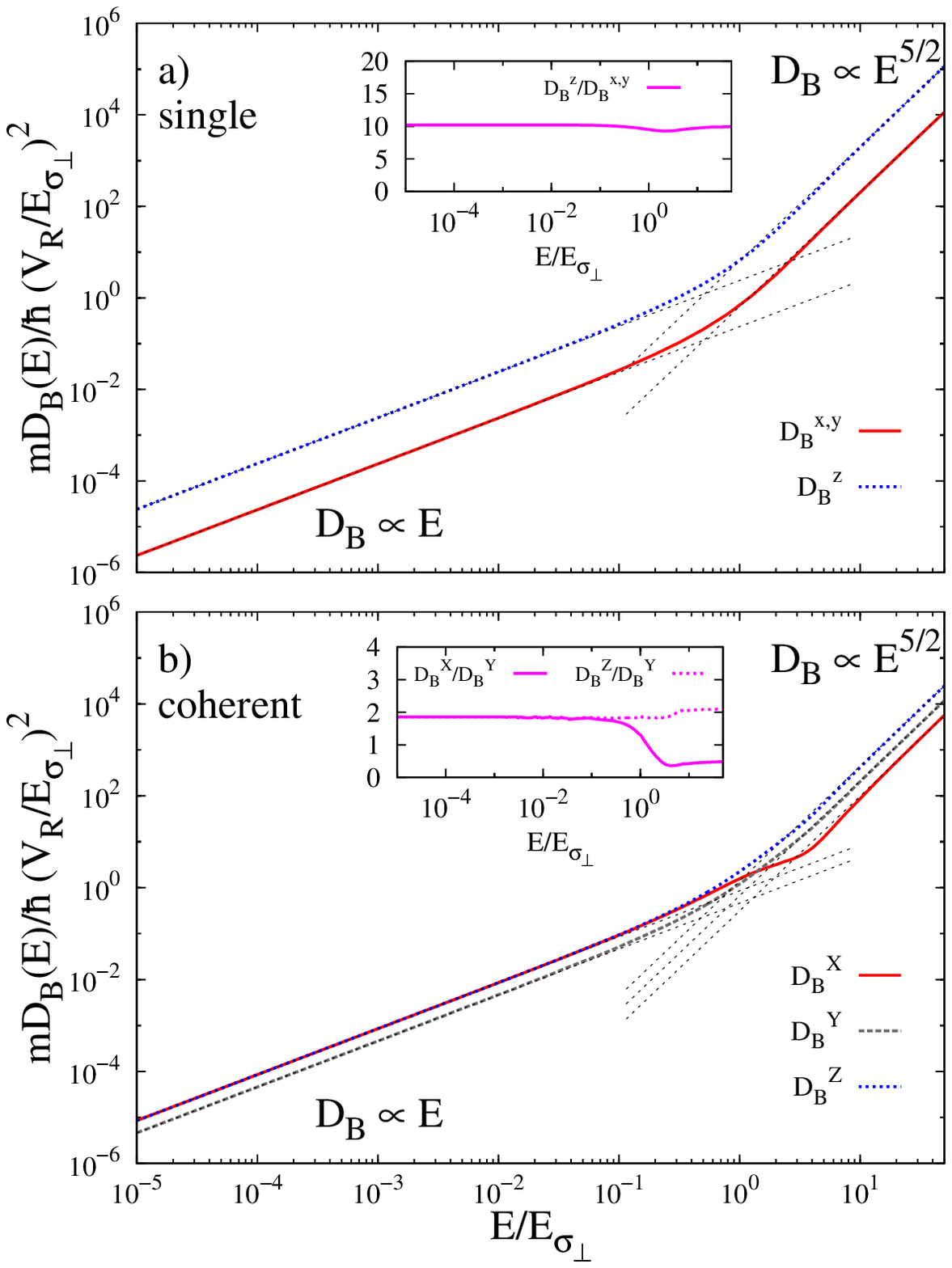}
\end{center}
\vspace{-0.6cm}
\caption{\small{(color online)
Boltzmann diffusion coefficients along the transport eigenaxes (eigencomponents of $\DiffTensB$) for the (a) single-speckle and (b) coherent-speckles configurations and for the parameters of fig.~\ref{fig:Speckle}.
The dotted lines are power-law fits ($D_\textrm{B}^u \propto E^{\gamma_u}$) to the data in the low and high energy limits.
The insets show the transport anisotropy factors.}}
\label{Figure:Boltzmann} 
\end{figure}

Figure~\ref{Figure:Boltzmann} shows the components of the Boltzmann diffusion tensor $\DiffTensB (E)$ versus the particle energy $E$, for the two above models of disorder.
Let us discuss first the single-speckle case, which not only forms a useful basis for the more complex coherent-speckles case but also shows several interesting properties itself.

Firstly, we find that the scattering time is shorter along the symmetry axis $z$ ($\tau_{\uveck_{z},E}<\tau_{\uveck_{\{x,y\}},E}$) for all values of $E$.
This property is simply explained by the fact that $\TFCor(\veck)$ has a wider extension in the plane $(k_x,k_y)$ orthogonal to $k_z$ than along $k_z$, which offers more scattering channels to particles travelling along $z$.
However, as it is well-known, the scattering and transport mean free times are different quantities in correlated disorder, due to angle-dependent scattering~\cite{abrikosov1975,chapman1991,rammer1998}.
In particular, the anisotropy of $\DiffTensB$ can significantly differ from that of $\tau_{\uveck,E}$.
Here, we find that the anisotropy of $\DiffTensB$ is reversed with respect to that of $\tau_{\uveck,E}$, so that $\DB^z > \DB^{x,y}$, for all values of $E$.It is due to the fact that the orbitals $\phi_{E,\uveck}^n$ contributing to $\DB^z$ in eq.~(\ref{DBE}) are associated to values of $\lambda_E^n$ significantly larger than those contributing to $\DB^{x,y}$.

Secondly, the low-energy behaviour of $\DiffTensB(E)$ does not reduce to a white-noise limit. It is due to the strong anisotropic divergence of $\TFCor (\veck)$ in the $|\veck| \rightarrow 0$ limit:
for $\vert\veck\vert \ll \sigma_{\perp}^{-1}$, we have 
$\TFCor (\veck) \sim \TFcor (\uveck)/\vert\veck\vert$
with $\TFcor(\uveck) = \exp[-(\sigma_{\parallel}/2\sigma_{\perp})^2 \hat{k}_z^2/(\hat{k}_x^2+\hat{k}_y^2)]/(\hat{k}_x^2+\hat{k}_y^2)^{1/2}$.
Using the fact that $\TFcor (\uveck)$ does not depend on $|\veck|$, we find that $\tau_{\uveck,E}$ and $\lambda^n_E$ do not depend on $E$, and that
  $\phi_{E,\uveck}^n$ is of the form $\varphi^n (\uveck)/\sqrt{\kE}$.
Then, all terms in eq.~(\ref{DBE}) are topologically unchanged and scale as $E$. The anisotropy of $\DiffTensB$ thus persists and is constant down to arbitrary low values of $E$, and $\DB^i \propto E$, as observed in fig.~\ref{Figure:Boltzmann}
for $\kE \ll \sigmaOrth^{-1}$ (\ie\ $E \ll E_{\sigmaOrth}$).
Conversely, a white-noise limit would lead to an isotropic diffusion tensor and $\DiffTensB \propto \sqrt{E}$.
In other words, in our case, the long-range disorder correlations lead to the anisotropic suppression of the white-noise limit.

Thirdly, in order to get further insight, it is useful to note that, for isotropic systems (see also refs.~\cite{kuhn2007,skipetrov2008,yedjour2010}), eq.~(\ref{eigeq}) is solved by the spherical harmonics $Y_l^m$, and that only the first term plus the $Y_1^m$ ($p$-level) harmonics contribute to $\DiffTensB$ in eq.~(\ref{DBE}).
For the anisotropic disorder we are considering, we find that the orbitals $\phi_{E,\uveck}^n$ are topologically similar to the spherical harmonics, \ie\ they show similar nodal surfaces~\footnote{Note that, in contrast to the isotropic case, here the $\lambda^{n}_E$ are not degenerated in a given $l$-like level.}.
For $\kE \ll \sigmaOrth^{-1}$, we find that $\DB^{x,y}$ is dominated by the first term in eq.~(\ref{DBE}) and $\DB^z$ by the $Y_1^0$-like orbital.
For $\kE \gg \sigmaOrth^{-1}$, the situation changes:
while $\DB^z$ is still dominated by the $Y_1^0$-like orbital, $\DB^{x,y}$ is now dominated by the $Y_1^{\pm 1}$-like orbitals with a contribution of the $Y_3^{\pm 1}$-like orbitals increasing with $E$.
Then, we find $\tau_{\uveck,E} \propto \kE$, and assuming weak topological change of the orbitals and the scaling $1-\lambda_E^n \propto 1/E$ (both confirmed numerically),
we get $\phi_{E,\uveck}^n \propto 1/\kE$ and $\DB^u(E) \propto E^{5/2}$,
as observed in fig.~\ref{Figure:Boltzmann} for $E \gg E_{\sigma_\perp}$.
This scaling was also found for isotropic speckle disorder~\cite{kuhn2007}.
Remarkably, in spite of the different contributing terms in eq.~(\ref{DBE}) at low and high values of $E$, the transport anisotropy is nearly independent of $E$
[see inset of fig.~\ref{Figure:Boltzmann}(a)].

Let us turn to the coherent-speckles case [fig.~\ref{Figure:Boltzmann}(b)], which is even richer owing to the more complex structure of the correlation function.
Note first that we recover the same general properties as for the single-speckle case, in particular the reversed anisotropies of scattering and diffusion, and the anisotropic suppression of the white-noise limit.
Here however, the transport eigenaxes are the bisectors
  $\{\uvecX,\uvecZ\}=(\uvecx\mp\uvecz)/\sqrt{2}$ and the axis
  $\uvecY=\uvecy$,
and the anisotropy is significantly smaller.
For $2\kE \lesssim 3.8\sigmaOrth^{-1}$, the behavior of $\DiffTensB (E)$ is governed by the central structure of $\tilde{\Cor}(\veck)$ since, in the on-shell Born approximation, a particle of energy $E$ probes $\TFCor(\veck)$ inside the $\veck$-space sphere of radius $2\kE$. Then, the directions $\uvecX$ and $\uvecZ$ are nearly identical but differ from the direction $\uvecY$ and we find $\DB^Y < \DB^X \simeq \DB^Z$.
The most interesting effect appears for $2\kE \gtrsim 3.8\sigmaOrth^{-1}$, \ie\ when the bumps of $\TFCor(\veck)$, 
located at $\veck \simeq \pm 3.8\sigmaOrth^{-1}\hat{\veck}_X$ for the parameters of fig.~\ref{fig:Speckle}(b), contribute to the transport.
The scattering time $\tau_{E,\uveck}$ becomes highly anisotropic and the orbital dominating $\DB^X$ is strongly distorted. As a result, $\DB^X$ is reduced and the corresponding anisotropy factor drops by a factor of $\simeq 4$. This effect happens to be strong enough to lead to the inversion of the transport anisotropy and we find $\DB^X<\DB^Y<\DB^Z$ for 
$2\kE \gtrsim 3.8\sigmaOrth^{-1}$ [see inset of fig.~\ref{Figure:Boltzmann}(b)].

\section{Localization}
\begin{figure*}[!t]
\begin{center}
\includegraphics[scale=0.5]{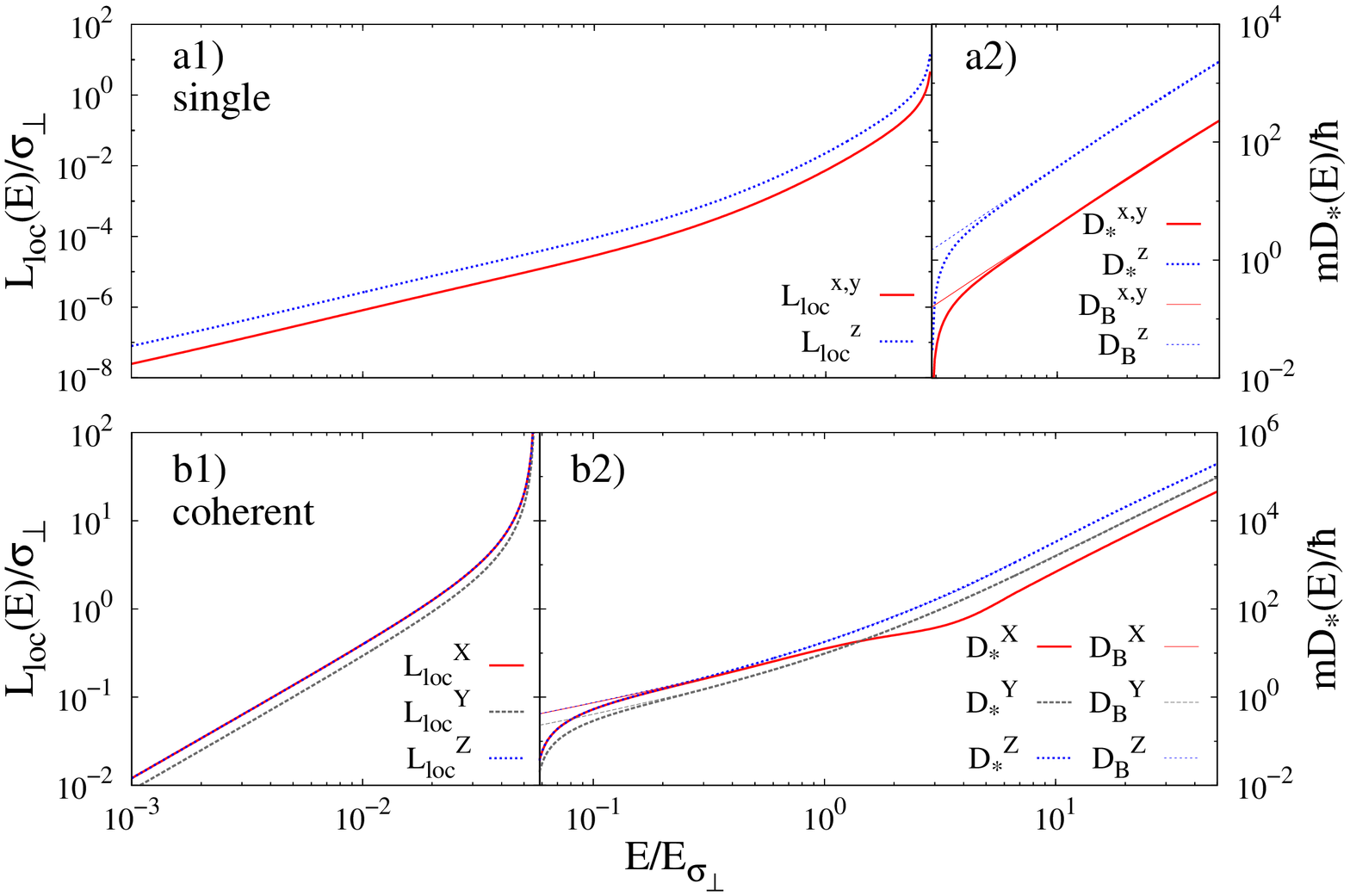}
\end{center}
\vspace{-0.6cm}
\caption{\small{(color online)
Components of $\LocTens$ ($E<\Emob$) and $\DiffTens$ ($E>\Emob$)
for the parameters of refs.~\cite{kondov2011,jendrzejewski2011} and fig.~\ref{fig:Speckle},
in the single-speckle (upper row; $\Vr=7.1E_{\sigmaOrth}$) and coherent-speckles (lower row; $\Vr=0.35E_{\sigmaOrth}$) cases. 
The components of $\DiffTensB$
are plotted for comparison (thin lines on the right column).
}}
\label{Figure:Loc} 
\end{figure*} 
So far, we have discussed incoherent Boltzmann diffusion.
We now include quantum corrections due to interference of the diffusing paths by incorporating the maximally-crossed diagrams (Cooperon and Hikami terms).
It yields the dynamic diffusion tensor $\DiffTens(\omega,E)=\DiffTensB(E)+\DiffTensCor(\omega,E)$ with~\cite{woelfe1984}
\be \label{weakloc}
\DiffTensCor(\omega,E)=
\frac{- \DiffTensB(E)}{\pi\hbar N_0(E)}
\int \! \frac{\ud \vecq}{(2 \pi)^3} \frac{1}{-i\omega + \vecq \! \cdot \! \DiffTensB(E) \! \cdot \! \vecq}.
\ee
As usual in such approach, the integral in eq.~(\ref{weakloc}) shows an ultraviolet divergence, which should be regularized.
To do so it is usefull to notice that eq.~(\ref{weakloc}) only depends on $\DiffTensB(E)$ [\ie\ in particular, not on $\TFCor(\veck)$], so that the anisotropy of the quantum corrections $\DiffTensCor$ is the same as that of $\DiffTensB$~\cite{woelfe1984}, and the only characteristic length in the direction $i$ is the transport mean free path $l_\textrm{B}^i (E) \equiv 3 \sqrt{m/2E}D_\textrm{B}^i(E)$.
This suggests an ellipsoidal cut-off of radius $1/l_\textrm{B}^i(E)$ along each transport eigenaxis, which we do here~\footnote{Although somewhat arbitrary the factor unity between the cut-off radius and $1/l_\textrm{B}^i(E)$ is justified by the agreement we find with another approach in the isotropic case, provided that the real part of the self energy is included, as discussed below.}.
It corresponds to an isotropic cut-off in the space rescaled according to the anisotropy factors of $\DiffTensB(E)$.
Then, following the standard self-consistent theory~\cite{vollhardt1992}, the above equations are solved for $\DiffTens(\omega,E)$ after replacing $\DiffTensB(E)$ by $\DiffTens(\omega,E)$ in the integrand of eq.~(\ref{weakloc}).
Hence, as is natural in scaling theory and also found with the sigma model for localization~\cite{woelfe1984, efetov1980}, the anisotropy of $\DiffTens(\omega,E)$ is the same as that of $\DiffTensB(E)$ by construction.
Proceeding (in 3D) in the long time limit ($\omega \rightarrow 0$),
a threshold energy $\Emob$ appears, solution of $\geomav{D_\textrm{B}}(\Emob) \equiv \det\{\DiffTensB(\Emob)\}^{1/3}=\hbar/\sqrt{3\pi}m$.
For $E>\Emob$, $\DiffTens(\omega,E)$ converges to a real definite positive tensor when $\omega \rightarrow 0$. It describes anisotropic normal diffusion, characterized by the propagation kernel~(\ref{BSEsingular}) where $\DiffTensB(E)$ is replaced by the quantum-corrected diffusion tensor $\DiffTens (E) \equiv \lim_{\omega \rightarrow 0}\DiffTens(\omega,E)$.
For $E<\Emob$, one finds $\DiffTens(\omega,E) \sim -i\omega \LocTensTwo(E)$ for $\omega \rightarrow 0$, where $\LocTensTwo(E)$ is a real positive definite tensor.
It characterizes exponential localization in the propagation kernel~(\ref{BSEsingular}), with the anisotropic localization tensor $\LocTens (E) \equiv \sqrt{\LocTensTwo (E)}$.

Figure~\ref{Figure:Loc} shows the components of $\LocTens$ (for $E<\Emob$) and $\DiffTens$ (for $E>\Emob$) for the single-speckle and coherent-speckles cases.
Due to the preservation of the transport anisotropy by the quantum corrections and the self-consistent procedure, the behavior of $\LocTens$ and $\DiffTens$ is completely determined by that of $\DiffTensB$.
The anisotropy factors of $\LocTens$ and $\DiffTens$ are nearly independent of $E$, except for the inversion of anisotropy of the coherent-speckles case.
For $E<\Emob$, the anisotropy factors of $\LocTens(E)$ are the square roots of those of $\DiffTensB (E)$.
In the low $E$ limit, we find $L_{\textrm{loc}}^i(E) \propto \big(D_\textrm{B}^i/\geomav{D_\textrm{B}}\big)^{1/2} E^{3/2}$.
When $E$ increases, $L_{\textrm{loc}}^i(E)$ grows and finally diverges at $\Emob$.
For $E>\Emob$, the anisotropy factors of $\DiffTens (E)$ are
the same as those of $\DiffTensB (E)$.
The quantum corrections are significant only close to $\Emob$.
For higher values of $E$,
$\DiffTens (E) \simeq \DiffTensB (E)$,
and in the high $E$ limit, we have $D_*^i(E) \propto (D_\textrm{B}^i/\geomav{D_\textrm{B}}) E^{5/2}$.

\section{Mobility edge}
The on-shell approach used so far
is expected to fairly describe the quantum transport properties~\cite{woelfe1984,vollhardt1992,kuhn2007}. It is however limited for quantitative estimates of the mobility edge, which require precise knowledge of the particle self energy in the disordered medium.
To estimate the mobility edge, one could in principle use the more sophisticated approaches of refs.~\cite{kroha1990,skipetrov2008,yedjour2010},
but the fact that the kind of disorder we are interested in
is continuous with fine anisotropic structures makes them hardly practicable numerically.
In order to overcome this issue, we propose
an alternative method based on the assumption that the leading term missing in the on-shell approximation is the real part of the self energy,
$\Sigma^\prime (\veck,E) = \textrm{P} \int \frac{\ud \veck'}{(2\pi)^3} \,  \frac{\tilde{\Cor}(\veck - \veck')}{E - \epsilon_{\veck^\prime}}$
(with $\textrm{P}$ the Cauchy principal value), which produces a shift of the energy states.
We incorporate $\Sigma^\prime (\veck,E)$ into the theory by averaging,
in first approximation, its $\veck$-angle dependence.
It amounts to replace the on-shell prescription by $\eps{\veck} = E^\prime \equiv E - \Delta (E)$ with
\be \label{eq:shift}
\Delta (E) \equiv \frac{1}{4\pi} \int_{\eps{\veck} = E -\Delta (E)} \ud\Omega_{\uveck}\ \Sigma^\prime \big(\veck,E\big)
\ee
where $\Omega_{\uveck}$ is the $\veck$-space solid angle.
Within this approach, all previous quantities are now regarded as functions of $E^\prime$ instead of $E$.
In particular, the mobility edge $\Emob$ is the solution of $\Emob - \Delta (\Emob) = \EmobNoShift$,
where $\EmobNoShift$ is determined using the on-shell approach.
Here two remarks are in order.
First, when applied to isotropic disorder, our method gives good agreement (within $5-7\%$) with the values of $\Emob$ found in ref.~\cite{yedjour2010} by a different approach, which in particular does not require the introduction of an ultraviolet cut-off.
Second, the $\veck$-angle averaging used in eq.~(\ref{eq:shift}) is justified {\it a posteriori} by the weak $\veck$-angle variations of $\Sigma^\prime$ around its average value at the mobility edge $\Emob$ (with standard deviations less than $10-15\%$).

\begin{figure}[t!]
\begin{center}
\includegraphics[scale=0.65]{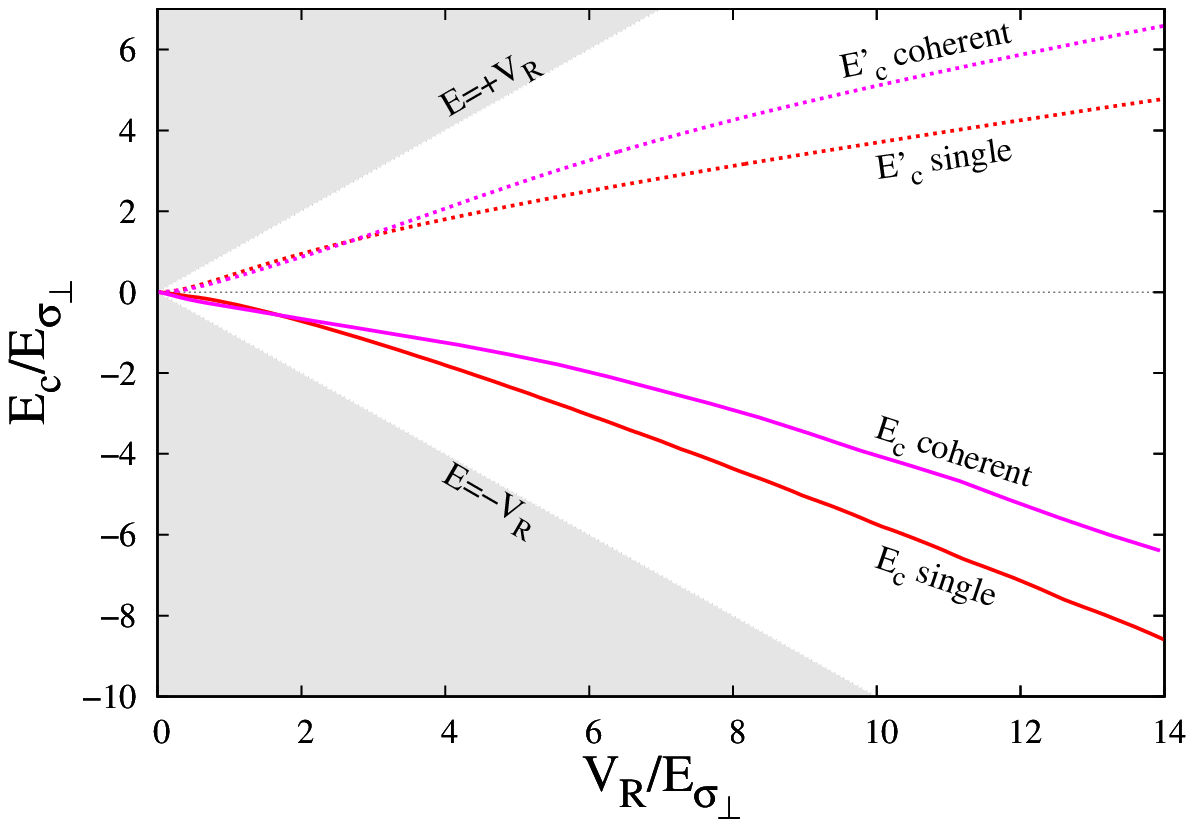}
\end{center}
\vspace{-0.6cm}
\caption{\small{(color online)
On-shell ($\EmobNoShift$) and corrected ($\Emob$) mobility edges
versus the disorder amplitude $\Vr$ for the single-speckle and coherent-speckles cases and the parameters of fig.~\ref{fig:Speckle}.}
}
\label{Figure:Emob} 
\end{figure}

For the single-speckle and coherent-speckles cases, the on-shell ($\EmobNoShift$) and corrected ($\Emob$) mobility edges are shown in fig.~\ref{Figure:Emob}. 
It is eye-catching that the shift of the energy states completely changes the behavior of the mobility edge.
While $\EmobNoShift$ is positive (\ie\ above $\av{V}$) and increases with $\Vr$, we find that $\Emob$ is negative (\ie\ below $\av{V}$) and decreases
with $\Vr$.
For $\Vr \lesssim E_{\sigmaOrth}$, this behavior is qualitatively similar to that found for isotropic disorder~\cite{yedjour2010}.
For larger values of $\Vr$, $\Emob$ further decreases, consistently with the expected behavior that $\Emob$ should approach the percolation threshold deep in the classical disorder regime ($\Vr \gg E_{\sigmaOrth}$)~\cite{shklovskii2008}.

\section{Discussion}
By considering matter waves in two particular models of optical disorder, we have shown
that structured disorder correlations in 3D can induce rich quantum transport and AL properties,
\eg\ 
  anisotropic suppression of the white-noise limit, 
  and inversion of the transport anisotropy with energy.
Such effects can be expected for other kinds of waves and/or other models of disorder,
and are particularly relevant to new systems where the disorder correlations can be controlled~\cite{kuhl2000,kuhl2008,barthelemy2008,lsp2010,clement2006,kondov2011,jendrzejewski2011}.
In the case of ultracold atoms, to which our study directly applies,
the transport properties can be probed by direct imaging of the atoms and control of the energy distribution.
This control does not need to be very fine.
For instance, in the single-speckle case, we found almost constant anisotropy factors ($D_*^{z}/D_*^{x,y} \simeq 10$ and $L_{\textrm{loc}}^{z}/L_{\textrm{loc}}^{x,y} \simeq 3.2$), and experimental data can be compared to these predictions almost independently of the energy distribution.
Although no precise value has been extracted from the experiment of ref.~\cite{kondov2011}, the experimental data indicate significantly larger anisotropy.
Further analysis would be required to clarify the origin of such a discrepancy.
Conversely, in the coherent-speckles case, the anisotropy factor was shown to be in fair quantitative agreement with theory~\cite{jendrzejewski2011}. The inversion of the transport anisotropy was however not observed
because the images were taken in the $(y,z)$ plane. It only gave access to $D^y=D^Y$ and $D^z=(D^X+D^Z)/2$, which do not show the inversion. In order to observe it, it is required to image the atoms along the transport eigenaxes and to tune the balance between the populations of low- and high-energy states.

Another major challenge is the evaluation of the mobility edge in 3D disorder with structured correlations. In this case numerical evaluations in advanced approaches~\cite{kroha1990,yedjour2010} are hardly practicable.
Here we have proposed and used an applicable approach.
In particular, it yields a mobility edge that is negative (\ie\ below $\av{V}$), as also predicted for isotropic speckle potentials~\cite{yedjour2010}.
Comparing to ref.~\cite{kondov2011}, our calculations hence significantly differ
from experimental values (\eg\ for $\Vr = 600 \,\textrm{nK}\times\kB \simeq 7.1 E_{\sigmaOrth}$, we find $\Emob \simeq -300\,\textrm{nK}\times\kB$ while $+900\,\textrm{nK}\times\kB$ is measured).
However, the method used in ref.~\cite{kondov2011} to infer $\Emob$ from the localized fraction uses the free-space kinetic energy distribution, which neglects the disorder-induced distortion of the energy distribution. It is questionable because the latter is, in particular, necessary to account for negative energy states (\ie\ below the disorder average value).
Comparing to ref.~\cite{jendrzejewski2011}, we find that $\Delta (\Emob)$ as calculated here is of the same order of magnitude as the heuristic shift introduced in ref.~\cite{jendrzejewski2011}
(\eg\ for $\Vr = h \times 680\,\textrm{Hz} \simeq 0.35E_{\sigmaOrth}$, we find $\Delta (\Emob)/h = -390\,\textrm{Hz}$ and the heuristic shift is $-225\,\textrm{Hz}$).
Given systematic uncertainties in the analysis of the experimental data, it is a fair agreement.
A precise test of the present theory would require a reliable determination of the energy distribution in ultracold-atom experiments, which is not available so far.

Beyond these direct applications, our results pave the way to further studies of anisotropy effects in coherent transport and AL. On the one hand, even more complex correlations can be designed in ultracold-atom experiments in 3D~\cite{piraud2012b} as well as in 2D~\cite{mrsv2010,pezze2011b}. On the other hand, it would be interesting to explore counterparts of the discussed effects for waves with different dispersion relations and/or in other kinds of controlled disorder~\cite{kuhl2000,kuhl2008,barthelemy2008}.
>From a theoretical point of view, several extensions of our approach may be envisioned. On the one hand, it would be worth going beyond the shifted on-shell approach used here, and incorporate the full structure of the particle spectral function. On the other hand, it would be interesting to estimate possible corrections to the anisotropy factors in the localized regime, which may be expected to be significant in strongly anisotropic disorder.
This would require developing an approach where the preservation of the Boltzmann transport anisotropy in the quantum corrected tensor is not built in the theory, as in the self-consistent approach used here.

\acknowledgments
We thank B.~DeMarco, F.~Jendrzejewski, V.~Josse, and B.~van~Tiggelen for useful discussions.
This research was supported by ERC (contract No.\ 256294).
The numerical calculations were performed on the LUMAT cluster (FR LUMAT 2764)
with the assistance of M.~Besbes.

\bigskip
APPENDIX
\section{Correlation functions}
In order to calculate the autocorrelation function of the disorder,
$\Cor(\vecr) \equiv \av{V(\vecr+\vecr^\prime)V(\vecr^\prime)}$,
for the single-speckle and coherent-speckles cases,
we use the paraxial approximation, which yields explicit formulas.
For the single-speckle case, we find
$\Cor(\vecr)=V_{\textrm{R}}^2\cor_{\textrm{1sp}}(x,y,z)$
with
\be
\cor_{\textrm{1sp}}(x,y,z) = 
\frac{1}{1+{4}z^2/\sigmaPara^2} \exp{\left[ - \frac{(x^2+y^2)/\sigmaOrth^2}{1+{4}z^2/\sigmaPara^2} \right]}.
\nonumber
\ee
For the coherent-speckles case, we find
$\Cor(\vecr)=(V_{\textrm{R}}/2)^2 \times \{\cor_{\textrm{1sp}}(x,y,z) + \cor_{\textrm{1sp}}(z,y,x) + 2\cor_{\textrm{coh}}(x,y,z)\}$
with
\beq
\cor_{\textrm{coh}}(\vecr) & = & \sqrt{\cor_{\textrm{1sp}} (x,y,z) \times \cor_{\textrm{1sp}} (z,y,x)}
\nonumber \\
& & \times \frac{(1+{{4}}\frac{xz}{\sigmaPara^2})\cos[\phi(\vect{r})]+{{2}}\frac{x-z}{{\sigmaPara}}\sin[\phi(\vect{r})]}
{\sqrt{1+{{4}}z^2/\sigmaPara^2} \sqrt{1+{{4}}x^2/\sigmaPara^2}}
\nonumber
\eeq
and
$\phi(\vect{r})=\frac{2\pi}{\lambda_{\textrm{L}}}(x-z)
-\frac{z}{{{\sigmaOrth^2\sigmaPara}}}{\frac{x^2+y^2}{1+{{4}}z^2/\sigmaPara^2}}
-\frac{x}{{{\sigmaOrth^2\sigmaPara}}}{\frac{z^2+y^2}{1+{{4}}x^2/\sigmaPara^2}}$.

\bibliographystyle{eplbib}
\renewcommand{\refname}{\uppercase{REFERENCES}}

\end{document}